# Machine Learning Based Cooperative Relay Selection in Virtual MIMO

Kunal Sankhe, Chandan Pradhan, Sumit Kumar and Garimella Rama Murthy
Signal Processing and Communication Research Center
International Institute of Information Technology, Hyderabad, India
Email: kunal.sankhe@research.iiit.ac.in, chandan.pradhan@research.iiit.ac.in, sumit.kumar@research.iiit.ac.in and rammurthy@iiit.ac.in

*Abstract*—In cellular systems, virtual multiple-input multiple-output (V-MIMO) technology promises to achieve performance gains comparable to conventional MIMO. In this paper, we propose cooperative relay selection algorithm based on machine learning techniques. Willingness of user to cooperate in V-MIMO depends on his current battery power, time and day along with incentives offered by service provider. Every user has different criterion to participate in V-MIMO, but follows a specific behavior pattern. Therefore, it is required to predict willing users in the neighborhood of source user (SU), before selecting users as cooperative nodes. Only inactive users belonging to Virtual Antenna Array (VAA) cell of SU are assumed to cooperate. This reduces control overheads in cooperative node discovery. In this paper, we employ prediction algorithm using two machine learning techniques i.e. ANN and SVM to find out inactive willing users within VAA cell. The parameters such as MSE, accuracy, precision and recall are calculated to evaluate performance of ANN and SVM model. Prediction using ANN has MSE of 3% with average accuracy of 97% (variance 0.37), whereas SVM has MSE of 2.58% with average accuracy of 97.56% (variance 0.17). We also observe that proposed prediction method reduces the node discovery time by approximately 29%.

## I. INTRODUCTION

Multiple Input Multiple Output (MIMO) is a key technology in 3GPP LTE and LTE-Advance standard [1]. MIMO can bring a number of potential benefits to mobile radio systems including more reliable operation in weak signal conditions, greater spectral efficiency and increased data rates for individual users. However, realization of MIMO in current cellular system is a big challenge due to limitations on size and complexity of user equipment (UE). To overcome this problem, V-MIMO can play a significant role, where inactive neighboring mobile stations (MSs) cooperate to form a virtual antenna array (VAA) with source user (SU) achieving spatial diversity, spatial multiplexing or beamforming. User cooperation is possible whenever there is at least a single MS willing to share the time-frequency resource at the cost of its own battery power. The main advantage of V-MIMO is the use of existing UE without the need for any extra infrastructure. It is particularly useful in scenarios with relatively high density of mobile stations to guarantee the availability of candidate cooperative nodes throughout the network. Higher the densities of mobile stations, higher the number of possible paths between a SU and eNodeB, which results in higher cooperative diversity gains. Additionally, V-MIMO helps to combat the dead spot problem which occurs in places with high shadowing and lack of line of sight such as subway train platforms, indoor environments and underground areas etc.

In a conventional cellular system, mobile users by default do not agree upon for cooperation, since it consumes scarce resources such as battery power, processor time and bandwidth. User may be ready to cooperate, only if some monetary benefits or other forms of incentives are provided by the service provider. Also, user's current battery power along with current time, day and location may also change his/her willingness decision. E.g. a particular user may be ready to cooperate in office duration, but not at peak hours. Also, one may be ready to cooperate only if battery power is above some threshold value. Sometimes incentive amount offered by the service provider may not be sufficient to persuade users to participate in cooperation. Also, it can happen that despite all cooperative and monetary benefits, few users may not be willing to participate in cooperation at any cost. Every user has different kind of willingness behavior which varies with day, time and location. But every user follows some specific pattern which can be learned over the time period and can be utilized to predict his willingness at current time and location. This information can be used before cooperative node discovery. In most of the previous works, it is assumed that all users are willing to participate in the cooperation. Only in [2], willingness status of the user is considered by default to be 0 and it changes to 1 if the user is ready to accept the incentive amount offered by the base station.

In this paper, we propose relay selection algorithm, which first predicts potential willing users located in the neighborhood of SU. The prediction of willing users will reduce cooperative node discovery time. The problem of prediction of willingness of the user is considered to be a binary classification problem. We have used machine learning techniques such as Artificial Neural Network (ANN) and Support Vector Machine (SVM) separately to classify whether the user is willing to cooperate or not. SU sends pseudorandom sequence to all these potential willing users. Users which are ready to cooperate will amplify and forward this message to eNodeB, whereas users who don't want to participate in V-MIMO will simply reject it. As end to end wireless link quality between SU and eNodeB via multiple cooperative nodes affects the performance of V-MIMO, eNodeB uses Bit Error Rate (BER) metric to select cooperative nodes.

The paper is organized into six sections. In section 3, we have briefly discussed the proposed algorithm. Section 4 deals with the description of the MLP NN architecture with Resilient Propagation learning algorithm and SVM classification technique. The experimental results are analyzed in Section 5. Section 6 concludes the paper.

## II. RELATED WORK

Different V-MIMO designs have been proposed in the literature to achieve spatial diversity [3], spatial multiplexing [4] and/or beamforming [5]. From relay perspective, amplify-and- forward [6], decode-and-forward [7] and compress-and-forward [8] protocols have been used to achieve communication in two-hops [7] or multi-hops [9]. Furthermore, V-MIMO can be studied at link level or system level [6]. Additionally, different metrics including ergodic capacity [6], outage capacity, bit error rate (BER) and energy efficiency have been analyzed to evaluate the performance of V-MIMO. The selection of relay users is also an important factor directly affecting the performance of V-MIMO. Several methods have been proposed for cooperative relay selection, some of the significant work is proposed in [10] where user pairing method based on orthogonality of the channel matrix is studied, whereas group-based user pairing is proposed in [11].

## III. PROPOSED ALGORITHM FOR RELAY SELECTION

In this section, we discuss the proposed algorithm for selection of cooperative users to form V-MIMO. We have considered stochastic geometry model to represent cellular V-MIMO network [12]. eNodeBs, inactive mobile users and source users are assumed to be spatially distributed in a Euclidean plane according to homogeneous Poisson point processes (PPPs). Any user can randomly become SU, who initiates a request to form V-MIMO. We assume that SU communicates with the nearest eNodeB and therefore, cell coverage area can be described as Voronoi tessellation [7] as shown by solid lines in Fig.1. Similarly, we assume that each inactive user can only cooperate with nearest SU. Therefore, each SU defines a VAA area called VAA cell as shown by dash lines in Fig.1. Voronoi tessellation serves a good model which captures the essence of random variation of SU and inactive users within the cell. Selection of cooperative nodes occurs in four steps. In first step, eNodeB filters inactive users based on the location of SU. eNodeB then predicts willingness of each user filtered from the previous step. In the third step, SU sends pseudorandom sequence to the predicted willing users, who amplify and forward it to eNodeB. Finally, eNodeB based on end to end link quality selects cooperative nodes having minimum BER. A stepwise relay selection algorithm is discussed below and also shown in Fig. 2.

### A. Location wise filtering of inactive users at eNodeB

SU who wants to initiate V-MIMO, sends request to eNodeB. Based on the location of SU, eNodeB filters out users which belong to the VAA cell of the corresponding SU. Since, only users belonging to VAA cell of SU are allowed to cooperate, interference caused by other SUs can be limited by appropriate transmit power control.

### B. Prediction of willingness of user nodes to cooperate

Using trained ANN engine or SVM model, eNodeB predicts the willingness of the users filtered from step A. ANN engine or SVM uses battery power of the user, current time, day and incentive amount offered by the service provider as features to predict the willingness of individual user. These features are important factors, which have direct impact on willingness decision of the user.

Users may participate in V-MIMO, only if battery power is well above the threshold value and this value may be different for each user. Similarly, some users may participate only during a particular time period. E.g. user may be ready to cooperate only during 12 - 3PM, but not during peak hours i.e. 5-7 PM. Additionally, users may have different behavior during weekdays than in weekends. Some users may ready to cooperate only if incentive amount given by the service provider is above certain threshold amount. Different categories of users such as household (HH), office users (OU), college students (ST) will have their own criteria to participate in V-MIMO as shown in Table I and show different behaviors towards cooperation.

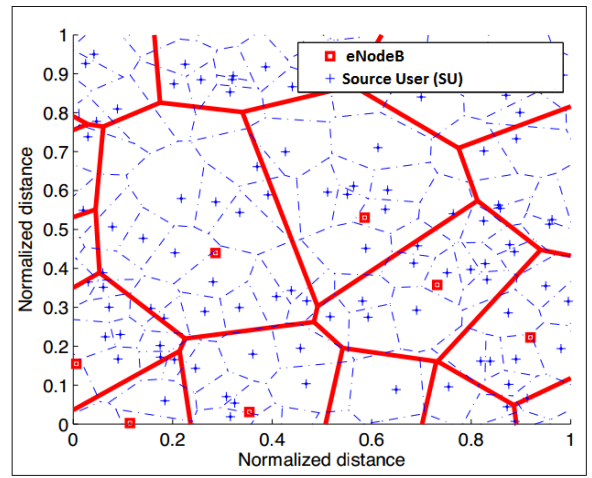

Fig. 1. Spatial structure of a cellular V-MIMO system with Voronoi eNodeB cells (solid lines) and Voronoi VAA cells (dashed lines) [7]

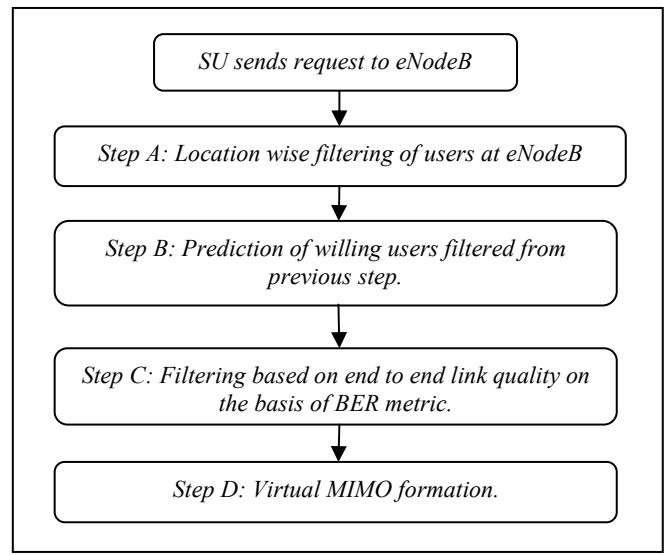

Fig. 2. Flow chart for relay selection algorithm

TABLE I. WILLINGNESS OF DIFFERENT TYPES OF USERS

| Users of different categories | Battery Power | Time | Day | Incentive Amount | Willing-ness |
|---|---|---|---|---|---|
| HH 1 | >0.3 | ~3 | Mon-Fri | >0.5 | Yes |
| HH 1 | <0.7 | 3,7,8 | Sat-Sun | <0.4 | No |
| OU 1 | >0.6 | ~3,9 | Mon-Sat | >0.6 | Yes |
| OU 1 | <0.5 | 7,8 | Sun | <0.8 | No |
| ST 2 | >0.3 | ~4,5 | Mon-Thu | >0.3 | Yes |
| ST 2 | >0.7 | ~3,7,8 | Fri-Sun | >0.7 | Yes |

a. *Battery Power & Incentive Amount is normalized between 0-1, Time is divided into 10 different zone*
b. *~ represents user not willing to cooperate in that particular time zone.*

Here, rule or policy based approach is not suitable as user's behavior may change after some duration. Therefore, eNodeB needs to learn the behavior of different users over a period of time and needs to adapt the model if user's behavior changes. Based on user's current battery power, time, day and incentive amount, eNodeB predicts willingness of the users filtered in the previous step and selects only those users for which willingness is true. Detailed explanation of prediction of willingness of individual user is discussed in section 4.

### C. Filtering based on link quality

It is necessary to determine the availability of the predicted willing users for cooperation before communication starts. To address this issue, we have used bit error rate (BER) metric to find out suitable cooperative relays which agree to participate in cooperation. SU sends a sequence of pseudorandom bits (known to eNodeB) to all the predicted willing nodes in a preselected broadcast channel. Nodes which are ready to cooperate will amplify the message and forward it to the eNodeB, whereas nodes which are not willing to cooperate will simply reject it. As the link between SU and each cooperative node as well as the link between each cooperative node and eNodeB both affects the performance of V-MIMO, the above method also helps to find out cooperative relays which provide good end to end path from SU to eNodeB. Now, eNodeB selects maximum $N-1$ cooperative nodes on the basis of BER, where $N$ is number of receiving antennas at eNodeB. As shown in Fig.3, the link SU-WU1-eNB has poor channel condition whereas SU-WU2-eNB provides good channel condition. Hence, eNodeB selects nodes which provide good end to end path with minimum BER. Later, eNodeB informs SU about selected willing nodes along with configuration parameters (e.g. transmit power, channel allocation, MIMO mode).

### D. Virtual MIMO Communication

The communication occurs in two phases: in the first phase, SU broadcasts information to selected cooperative nodes; in the second phase SU along with cooperative nodes send data to eNodeB simultaneously using distributed SFBC (Space Frequency Block Coding) [13]. As cooperative nodes are located nearer to SU, higher order modulation scheme can be used in first phase due to availability of good channel condition between SU and cooperative nodes. If no willing users are predicted in the step B, then SU directly communicates with eNodeB without any cooperative nodes.

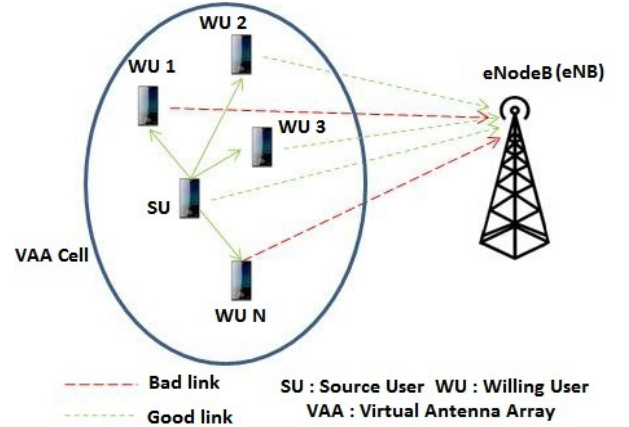

Fig. 3. Filtering based on link quality of end to end path.

## IV. PREDICTION FOR WILLINGNESS OF USERS

In this section, we discuss two machine learning techniques to predict the willingness of the users within a VAA cell of SU. Neural network based technique is discussed first, where multilayer feed-forward network is trained using Resilient Propagation learning algorithm. Second approach is based on Support Vector Machine (SVM), which is frequently used classification technique.

### A. ANN based approach

Artificial Neural Network (ANN) is a computational model inspired by human brain. It is a powerful tool to process problems having non-linear and complex data, even if data is imprecise and noisy[14]. ANN can identify and learn correlated patterns between input data and corresponding target values. After training, ANN can be used to predict the outcome of new independent input data. The general multilayer feed-forward neural network (NN) architecture is shown in the Fig. 4.

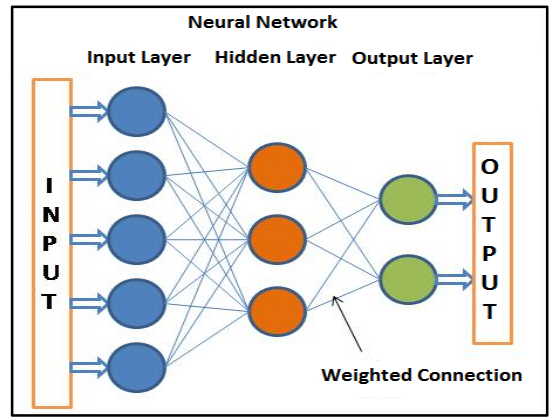

Fig. 4. Architecture of Multilayer Feed-forward Neural Network

For our study, we have used multilayer perceptron (MLP) NN to predict willingness of user to cooperate in V-MIMO. It consists of one hidden layer and one output layer. The

activation functions of all these layers are 'tanh' function which is given as:

$$f(x) = \tanh(x) = \frac{e^x - e^{-x}}{e^x + e^{-x}} \quad (1)$$

The Resilient Propagation (RPROP) [15] learning algorithm, which is an improved version of back-propagation, is used to update the weights. RPROP assumes that different weights need different step sizes for updates, which also vary throughout the learning process. The step size value depends on its local sight on the error function E and is given by following rule:

$$\Delta_{ij}(t) = \begin{cases} \eta^+ * \Delta_{ij}(t-1), & if \frac{\partial E(t-1)}{\partial w_{ij}}\frac{\partial E(t)}{\partial w_{ij}} > 0 \\ \eta^- * \Delta_{i,j}(t-1), & if \frac{\partial E(t-1)}{\partial w_{ij}}\frac{\partial E(t)}{\partial w_{ij}} < 0 \\ \Delta_{ij}(t-1), & otherwise \end{cases} \quad (2)$$

where $0 < \eta^- < 1 < \eta^+$

If the error gradient for a given weight $w_{ij}$ has the same sign in two consecutive epochs, its step size $\Delta_{ij}$ is increased. On the other hand, if the sign is switched, the step size is decreased. Weight updates are given as follows:

$$w_{i,j}(t) = \begin{cases} -\Delta_{i,j}(t), & if \frac{\partial E(t)}{\partial w_{i,j}} > 0 \\ +\Delta_{i,j}(t), & if \frac{\partial E(t)}{\partial w_{i,j}} < 0 \\ 0, & otherwise \end{cases} \quad (3)$$

The performance of the ANN model is evaluated in terms of mean squared error (MSE) between the predicted value and the target data for a selection of appropriate training set.

$$MSE = \frac{1}{N}\sum_{i=1}^{N}(y_i - d_i)^2 \quad (4)$$

where $y_i$ = output value calculated by the network, $d_i$ = expected output, $N$ = number of samples.

### B. SVM based approach

Support Vector Machines (SVMs) are among the best supervised learning models. In this subsection, we briefly discuss the use of SVM in classification problem. Detailed level of explanation is given in [16].

Let the training set of instance-label pairs be $(x_i, y_i)$, $i = 1,..l$ with each input $x_i \in R^n$ denotes a pattern to be classified and $y_i \in \{1, -1\}$ denotes class labels. SVM first maps training vectors $x_i$ into higher dimensional feature space $\mathcal{F}$ by function $z = \emptyset(x_i)$. The SVM constructs a hyperplane $w^Tz + b$ in $\mathcal{F}$ for which the separation between positive and negative examples is maximized. $w$ for this "optimal" hyperplane can be written as $w = \sum_{i=1}^{l}\alpha_i y_i \emptyset(x_i)$, where Lagrange multipliers $\alpha = (\alpha_1, ..., \alpha_l)$ can be found by solving the following Quadrature Programming (QP) problem [17]:

$$max\ W(\alpha_1, \alpha_2, ..., \alpha_l) = \sum_{i=1}^{l}\alpha_i - \frac{1}{2}\sum_{i=1}^{l}\sum_{j=1}^{l}\alpha_i\alpha_j y_i y_j K(x_i, x_j) \quad (5)$$

*subject to*

$$0 \leq \alpha_i \leq C, \quad i = 1,2,...,l$$
$$\sum_{i=1}^{l}\alpha_i y_i = 0$$

Mapping $\emptyset(.)$ from $R^n$ to $\mathcal{F}$ never appears explicitly, but it is necessary to define kernel function $K(x_i, x_j) \equiv \emptyset(x_i)^T\emptyset(x_j)$ which implicitly defines $\emptyset(.)$. E.g. kernel for radial basis function is $K(x_i, x_j) = \exp\left(-\gamma\|x_i - x_j\|^2\right), \gamma > 0$.

For those $\alpha$'s greater than zero, the corresponding training examples must lie along the margins of the decision boundary (by the Kuhn–Tucker theorem), and these are called the support vectors as shown in Fig. 5.

During testing, for a test vector $x \in R^n$, we first compute,

$$\alpha(x, w) = \sum_{i=1}^{l}\alpha_i y_i K(x, x_i) + b \quad (6)$$

and then its class label $o(x, w)$ is given by

$$o(x, w) = \begin{cases} 1, & \alpha(x, w) > 0 \\ -1, & otherwise \end{cases} \quad (7)$$

When the training set is not separable in $\mathcal{F}$, the SVM algorithm introduces nonnegative slack variable $\xi_i \geq 0; i = 1, ..., l$. The resultant problem becomes

$$\min_{w,b,\xi}\frac{1}{2}w^Tw + C\sum_{i=1}^{l}\xi_i \quad (8)$$
$$subject\ to\ y_i(w^T\emptyset(x_i) + b) \geq 1 - \xi_i$$
$$\xi_i \geq 0$$

Here, $C$ is user-defined regularization parameter controlling the trade-off between model complexity and training error, and $\xi_i$ measures the absolute difference between $\alpha(x_i, w)$ and $y_i$.

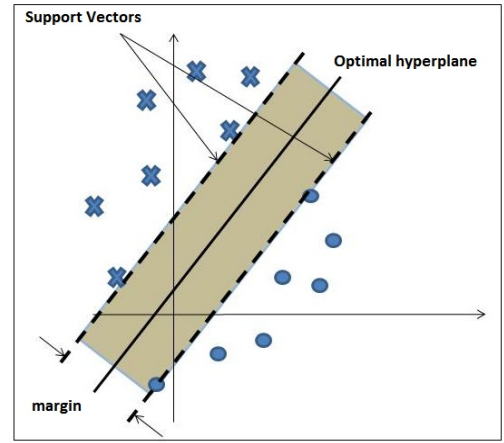

Fig. 5. SVM classification with a hyperplane that maximizes the separatingmargin between the two classes (indicated by data points marked by "x"s and"o"s).

## V. SIMULATIONS AND RESULTS

For our experiment, we have simulated a single cell scenario with users randomly located using two dimensional Poisson point processes. Few users are randomly selected as SUs who initiate the request to form V-MIMO. According to Voronoi tessellation, each SU has its own VAA cell within which inactive users can cooperate with SU in V-MIMO. We have randomly considered one SU among other SUs to verify our proposed algorithm. eNodeB predicts willingness of only those users which belong to VAA cell of the corresponding SU.

The objective of ANN based approach was to design a neural network model, which can predict user's willingness to cooperate in V-MIMO. We have used open source 'Encog' machine learning framework to implement multilayer feed-forward neural network [18]. The problem of prediction of willingness of the user is formulated as a binary classification problem where battery power, time, day and incentive amount offered by the service provider are used as features and willingness of the user is considered as a class label. Artificial dataset is created to verify the algorithm. Network is trained for individual user using resilient propagation (RPROP) as a learning algorithm. The training phase of the network continues by adaptively adjusting weights and step size, until MSE reaches to threshold value, which is 0.01 in our experiment. Encog library by itself scales the features to get appropriate results. Due to scaling, we have a network with 23 neurons in the input layer, 10 neurons in the hidden layer and two neurons in the output layer. To evaluate the performance of ANN model, along with (MSE), we have also calculated other classification metrics such as accuracy, precision and recall which are given as below:

$$Accuracy = \frac{TP + TN}{TP + FP + TN + FN}; Precision = \frac{TP}{TP + FP}$$

$$Recall = \frac{TN}{TN + FN}$$

where TP = number of *true positives* (number of times user is correctly predicted for his willingness to cooperate); FP = number of *false positives* (number of times user is incorrectly predicted for his willingness to cooperate); TN = number of *true negatives* (number of times user is correctly predicted for his unwillingness to cooperate); FN = number of *false negatives* (number of times user is incorrectly predicted for his unwillingness to cooperate).

We have evaluated the above metric parameters for various numbers of inactive users in a VAA cell of SU as shown in Table II.

To predict willingness of users using SVM, we have used '*LIBSVM*' library [19]. We have considered the same features as used in ANN. First data are transformed to the format supported by the SVM package. Then data are scaled using appropriate scaling method. For this experiment, we have used RBF kernel which nonlinearly maps samples into higher dimensional space. RBF kernel has two parameters $C, \gamma$ and optimal values of $C, \gamma$ are not known beforehand. In this experiment, we have used 3-fold cross-validation to find best value of $(C, \gamma)$, which improves the prediction accuracy.

TABLE II. PERFORMANCE METRIC OF MLP-RPROP NN

| No. of inactive users in a VAA cell | Mean Squared Error (MSE) (%) | Accuracy (%) | Precision (%) | Recall (%) |
|---|---|---|---|---|
| 10 | 3.95 | 96.05 | 85.32 | 90.00 |
| 15 | 3.33 | 96.67 | 88.37 | 90.28 |
| 20 | 2.82 | 97.18 | 92.40 | 92.87 |
| 25 | 2.58 | 97.42 | 93.65 | 94.27 |
| 30 | 2.32 | 97.70 | 94.40 | 94.52 |

Performance metrics similar to ANN are evaluated for SVM model and tabulated in Table III. MSE and other performance metrics of prediction using ANN model and SVM model are plotted in Fig. 6 & 7.

TABLE III. PERFORMANCE METRIC OF SVM

| No. of inactive users in a VAA cell | Mean Squared Error (MSE) (%) | Accuracy (%) | Precision (%) | Recall (%) |
|---|---|---|---|---|
| 10 | 3.20 | 96.80 | 87.96 | 91.94 |
| 15 | 2.60 | 97.40 | 90.02 | 93.52 |
| 20 | 2.17 | 97.82 | 93.55 | 95.96 |
| 25 | 2.14 | 97.86 | 94.10 | 95.96 |
| 30 | 2.05 | 97.92 | 94.72 | 96.02 |

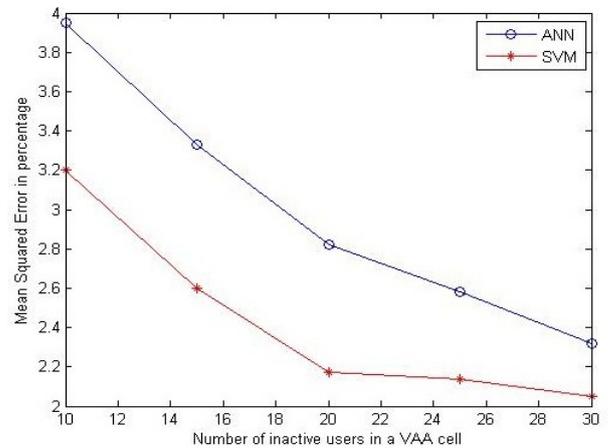

Fig. 6. MSE of prediction using MLP-RPROP Neural Network and SVM

After location filtering, SU broadcasts pseudorandom message within a VAA cell. Inactive nodes forward it to eNodeB using amplify-and-forward method for further filtering based on the link quality. eNodeB selects first $N - 1$ nodes as cooperative nodes which satisfies minimum BER requirement to reduce relay discovery time. Waiting time at eNodeB to get response from inactive users is directly proportional to number of users within VAA cell. Due to prediction, waiting time at eNodeB significantly reduces, as eNodeB has to wait for the duration proportional to predicted willing users. In Fig. 8, we have compared average time required by conventional algorithm with proposed algorithm assuming 50 inactive users.

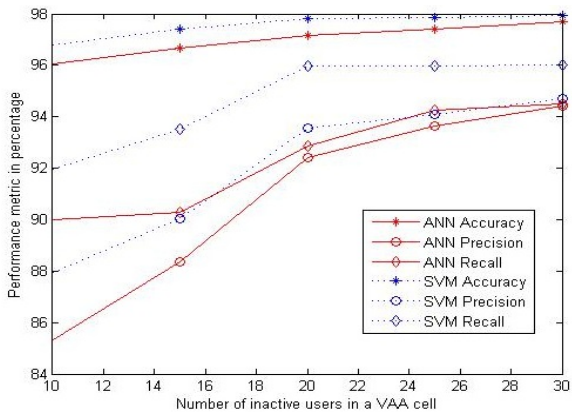

Fig. 7. Accuracy, precision and recall of prediction using MLP-RPROP Neural Network modeland SVM model

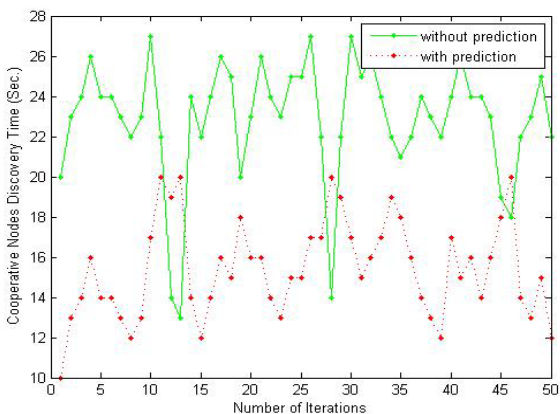

Fig. 8. Average time required for cooperative nodes discovery with and without prediction.

Based on number of cooperative nodes selected, eNodeB forms V-MIMO with SU using distributed SFBC coding.

## VI. CONCLUSION

We have proposed a cooperative relay selection algorithm based on machine learning techniques. Using idea of Voronoi tessellation, users located within VAA of the SU are only allowed to cooperate with SU. Thus, limiting the interference caused by other SUs. Applying ANN or SVM model, user willingness for cooperation is predicted. Additionally channel quality estimation is performed in order to select the nodes which possess better link quality between the SU and eNodeB. All the complex signal processing tasks occur at eNodeB, thus conserving energy of SU and other mobile users. Performance metric such as MSE, accuracy, precision and recall are computed to evaluate the performance of ANN and SVM model. We found that prediction based approach selects the relay nodes and creates the V-MIMO in a significantly lesser time compared to the conventional relay selection algorithm.